\documentclass[epj]{svjour}
\usepackage{graphics}
\usepackage{rotating}
\begin{document}
\title{The nature of the $\Lambda(1405)$}
\author{E. Oset\inst{1}, L.S. Geng\inst{1}\and M. D\"oring\inst{2} 
\thanks{\emph{Present address:} oset@ific.uv.es}%
}                     
\offprints{}          
\institute{ Departamento de F\'{\i}sica Te\'orica and IFIC, Centro
Mixto, Institutos de Investigaci\'on de Paterna - Universidad de
Valencia-CSIC, Valencia, Spain\and IKP, Institut f\"ur Kernphysik, 
Forschungszentrum J\"ulich,
  52425 J\"ulich, Germany}
\date{Received: date / Revised version: date}
%
\abstract{We present here some results supporting the nature of the
$\Lambda(1405)$ resonance as dynamically generated from the meson baryon
interaction in coupled channels and resulting from the superposition of two
close-by poles. We find support for this picture in the 
$K^- p \to \pi^0 \pi^0 \Sigma^0 $ reaction, which shows a different shape than
the one obtained from the $\pi ^- p \to K^0 \pi \Sigma $ reaction. We also call
the attention to the $K^- p \to \gamma \pi \Sigma$ with $\pi \Sigma$ in 
the $\Lambda(1405)$ region,
which shows a narrow peak in the calculations around 1420 MeV. We also report on
recent calculations of the radiative decay of the two $\Lambda(1405)$ states and
on reactions to obtain information on these decay modes.  Finally, we present
results for the  $pp\rightarrow p K^+\Lambda(1405)$ reaction recently measured
at ANKE/COSY and compare them with theoretical results.
\PACS{
      {PACS-key}{}   \and
      {PACS-key}{}
     } 
} 
\maketitle
\section{Introduction}
\label{intro}
The $\Lambda(1405)$ resonance has a long history as a dynamically generated
resonance from the interaction of meson baryon coupled channels
\cite{dalitz,jennings} but its study has become more systematic with the
advent of chiral unitary theories 
\cite{weise,Kaiser:1996js,angels,Oset:2001cn,Oller:2000fj,jido,Garcia-Recio:2002td,Garcia-Recio:2003ks,Hyodo:2002pk,GarciaRecio:2005hy}.
 One of the interesting
surprises was that in \cite{jido} it was found that the empirical
$\Lambda(1405)$ corresponded in fact to two poles (see fig. 1).

\begin{figure}
\resizebox{0.50\textwidth}{!}{%
  \includegraphics{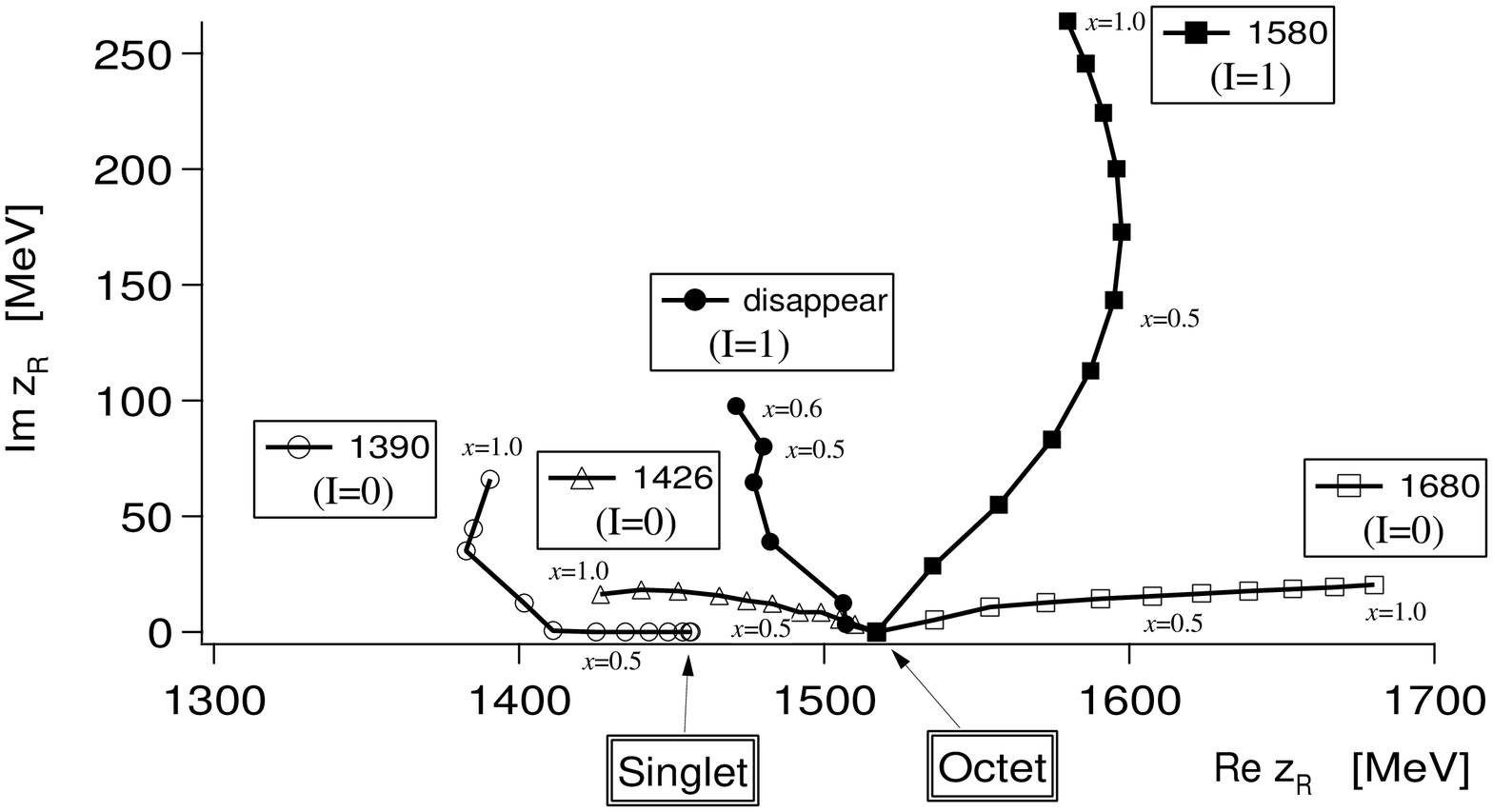}
}
\caption{Trajectories of the poles in the scattering amplitudes obtained by
  changing the SU(3)  breaking parameter $x$ gradually from zero to one. At 
  the SU(3) symmetric 
  limit ($x=0$),
   only two poles appear, one is for the singlet and the other for the octet.
  The symbols correspond to the step size $\delta x =0.1.$}
\label{fig:1}       
\end{figure}

In the figure one can see trajectories which are obtained when SU(3) symmetry is
broken gradually. Recall that we have the interaction of the octet of
pseudoscalar mesons and the octet of stable baryons. Hence 
\begin{equation}
 8 \otimes 8=1\oplus 8_s \oplus 8_a \oplus 10 \oplus \overline{10} \oplus 27~,
\end{equation}
and we get two octets and a singlet where the interaction is attractive. When
all the masses of mesons are made equal and the baryon masses equal (SU(3)
symmetric case) there is a degeneracy of the two octets. When the masses gradually
move to the physical ones the poles follow the trajectories in the figure and
then we see that in the region of the $\Lambda(1405)$ there are two poles one
coming from an octet and the other one from the singlet. The one around 1420 MeV
is narrow and couples mostly to $\bar{K}N$ while the one around 1395 MeV is wide
and couples mostly to $\pi \Sigma$. Recently there have been many works
implementing effects of higher order Lagrangians in $\bar{K}N$ and coupled
channels \cite{Borasoy:2005ie,Oller:2005ig,Oller:2006jw,borasoy}. 
They all share these basic
features but there are variations as to the width of the lighter state, which is
very wide in all approaches. The study of theoretical uncertainties in
\cite{borasoy} shows that the lowest order results of \cite{angels} fall within
theoretical uncertainties.

\section{The $K^- p \to \gamma \pi \Sigma$ in the $\Lambda(1405)$ region}
\label{sec:1}
With the discussion above one expects that different reactions produce different
shapes for the $\Lambda(1405)$, depending on which of the two states is more
strongly excited. In this sense, let us take the process
$K^- p \to \gamma \pi \Sigma$ with the $\pi \Sigma$ in the $\Lambda(1405)$ region. This reaction 
was studied
in \cite{nacher} and was found to be dominated by initial state radiation from
the $K^-$. Hence, the $\Lambda(1405)$ is formed from $K^- p$ with a $K^-$ that
has lost energy and we should expect it to be dominated by the high energy pole
that couples strongly to $\bar{K}N$. This is indeed the case as can be seen in
fig. 2. The peak of the cross section appears around 1420 MeV and the width is
about 30 MeV, narrower than the nominal one of around 50 MeV. This is one of the
experiments that we encourage to be performed and which will provide useful
information on the nature of the $\Lambda(1405)$ and its two pole structure. 

\begin{figure}
\begin{center}
\begin{turn}{-90}
\resizebox{0.30\textwidth}{!}{%
  \includegraphics{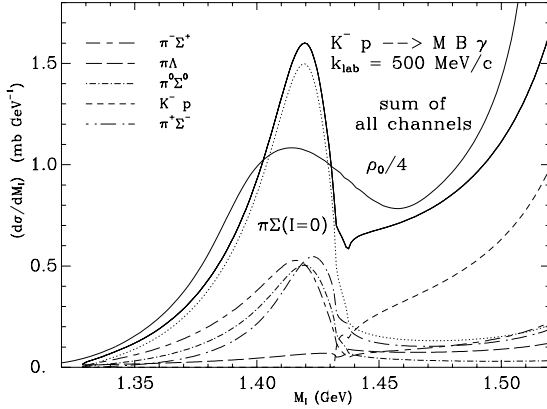}
}
\end{turn}
\caption{Mass distribution for the different channels, $\pi^+ \Sigma^-$,
$\pi^- \Sigma^+$, $\pi^0 \Sigma^0$, $\pi \Lambda$, $K^- p$. The solid line
with the resonance  shape is the sum of cross sections for all channels.
Dotted line: pure $I=0$ contribution from the $\Sigma\pi$ channels.
The effects of the Fermi motion with ($\rho=\rho_0/4$) is shown with solid
line. The labels for the other lines are shown in the figure.}
\label{fig:2}   
\end{center}
\end{figure}

\section{The $K^- p \to \pi^0 \pi^0 \Sigma^0 $ reaction}
 This reaction was studied experimentally in \cite{prakhov} and theoretically in
 \cite{magas}. The reaction is similar to the former one but the process is
 dominated by one $\pi^0$ emission from the initial proton, as can be seen in
 fig. 3. 
 
  \begin{figure}
\begin{center}
\resizebox{0.40\textwidth}{!}{%
  \includegraphics{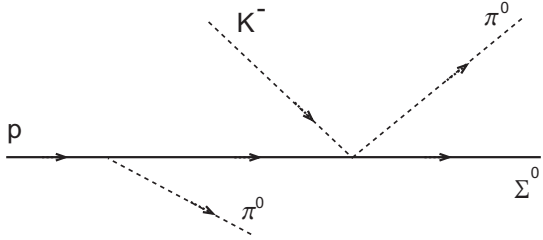}
}
\caption{Dominant mechanisms for the $K^- p \to \pi^0 \pi^0 \Sigma^0 $ reaction }
\label{fig:3}   
\end{center}
\end{figure}
 
  Once the $K^- p$
 system loses energy after the $\pi^0$ emission, we can have the right energy to
 produce the $\Lambda(1405)$ which decays into $\pi^0 \Sigma^0$. The interesting
 thing is that the $\Lambda(1405)$ is once again excited by $\bar{K}N$ and this 
 will
 give weight to the higher mass pole, resulting in a peak at higher masses and
 narrower than in other experiments which are dominated by the lighter mass
 pole. This is exactly what happens as one can see in fig. 4, where the shape of
 the $\Lambda(1405)$ for the  $K^- p \to \pi^0 \pi^0 \Sigma^0 $ reaction are
 superposed with that of the $\pi^- p \to K^0 \pi \Sigma$, which according to
 \cite{hyodo} is dominated by the smaller mass pole. 
 
 \begin{figure}
\begin{center}
\resizebox{0.40\textwidth}{!}{%
  \includegraphics{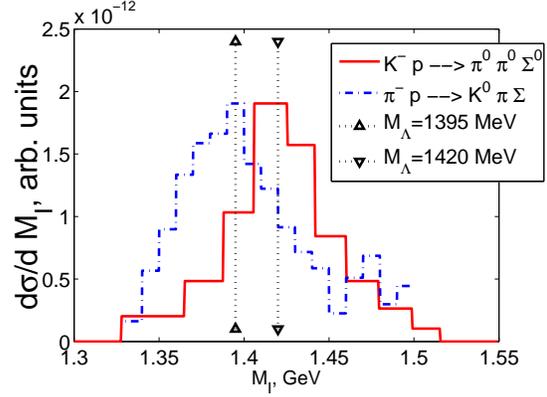}
}
\caption{Two experimental shapes of  $\Lambda(1405)$ resonance}
\label{fig:4}   
\end{center}
\end{figure}

\section{Radiative decay of the $\Lambda(1405)$ }
Radiative decay of the  $\Lambda(1405)$ can be another tool to learn about the
nature of this resonance. A recent study from the present perspective has been
done in \cite{gengmisha}. The idea is that since the resonance is like a
molecule of meson baryon  in different channels, the photon must be radiated from
the meson or baryon components. Then the determination of the radiative width is
done by evaluating the diagrams of fig. 5.
 \begin{figure}
\begin{center}
\resizebox{0.50\textwidth}{!}{%
  \includegraphics{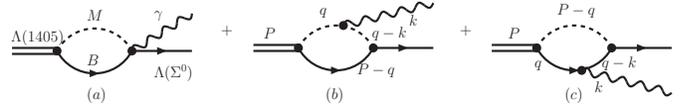}
}
\caption{The radiative decay mechanism of the
$\Lambda(1405)$, where $MB$ can be any of the four charged channels
of the ten coupled channels:
 $K^-p$, $\bar{K}^0n$,
$\pi^0\Lambda$, $\pi^0\Sigma^0$, $\eta\Lambda$, $\eta\Sigma^0$,
$\pi^+\Sigma^-$, $\pi^-\Sigma^+$, $K^+\Xi^-$, and $K^0\Xi^0$.}
\label{fig:5}   
\end{center}
\end{figure}

The results obtained are shown in table 1, where they are compared with other
models.  

\begin{table*}[htpb]
\renewcommand{\arraystretch}{1.5}
\setlength{\tabcolsep}{3mm}
 \centering \caption{\label{table:decaywidth}The radiative decay widths  of the $\Lambda(1405)$
 predicted by different theoretical models, in units of keV. The values denoted by ``U$\chi$PT'' are
 the results obtained in the present study. The widths calculated for the low-energy pole and high-energy
 pole are separated by a comma.} 
\begin{tabular}{c|ccccc}
\hline\hline
 Decay channel & U$\chi$PT& $\chi$QM \cite{Yu:2006sc} & BonnCQM \cite{VanCauteren:2005sm} &
NRQM & RCQM \cite{Warns:1990xi}\\\hline
 $\gamma\Lambda$ & $16.1,\,64.8$ & 168 & 912 & 143 \cite{Darewych:1983yw}, 200, 154 \cite{Kaxiras:1985zv} & 118 \\\
 $\gamma\Sigma^0$& $73.5,\,33.5$ & 103 & 233 & 91 \cite{Darewych:1983yw}, 72, 72 \cite{Kaxiras:1985zv}& 46 \\\hline\hline
Decay channel & MIT bag \cite{Kaxiras:1985zv} & chiral bag \cite{Umino:1992hi} & soliton~\cite{Schat:1994gm}&algebraic model~\cite{Bijker:2000gq}& isobar fit~\cite{Burkhardt:1991ms}\\\hline
 $\gamma\Lambda$ &60, 17 & 75 &44,40 & 116.9&$27\pm8$\\
$\gamma\Sigma^0$ &18, 2.7 & 1.9 &13,17& 155.7&$10\pm4$ or $23\pm7$\\
 \hline\hline
\end{tabular}
\end{table*}

As one can see in the table,  the values we obtain for the radiative decay into
$\gamma \Lambda$ and $\gamma \Sigma$ depend on which one is the state that
decays. This is novel to other approaches where there is only one
$\Lambda(1405)$. Obviously in different experiments one will get a different
combination of the contribution of the two poles. For this reason in
\cite{gengmisha} two reactions were studied which give different weight to either
of the poles, resulting in different shapes for the $\Lambda(1405)$ resonance
and different strength for the radiative width. We studied the reactions 
$K^-p\rightarrow \pi^0\gamma\Lambda(\Sigma^0)$ 
and $\pi^- p\rightarrow K^0\gamma\Lambda(\Sigma^0)$. We found different strength
and shapes for the two cases and the $\Lambda$ or $\Sigma$ in the final state.
The results can be seen in figs. 6 and 7. 

 \begin{figure}
\begin{center}
\resizebox{0.50\textwidth}{!}{%
  \includegraphics{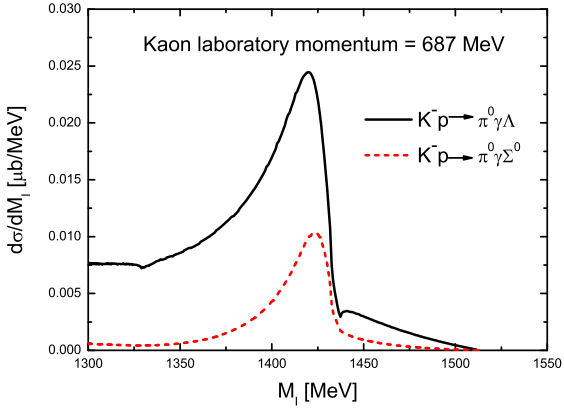}
}
\caption{The invariant mass distribution of $K^-p\rightarrow \pi^0\gamma
\Lambda(\gamma\Sigma^0)$ as a function of the invariant mass of the final
$\gamma\Lambda$($\gamma\Sigma^0$) system.}
\label{fig:6}   
\end{center}
\end{figure}

 \begin{figure}[htpb]
\begin{center}
\resizebox{0.50\textwidth}{!}{%
  \includegraphics{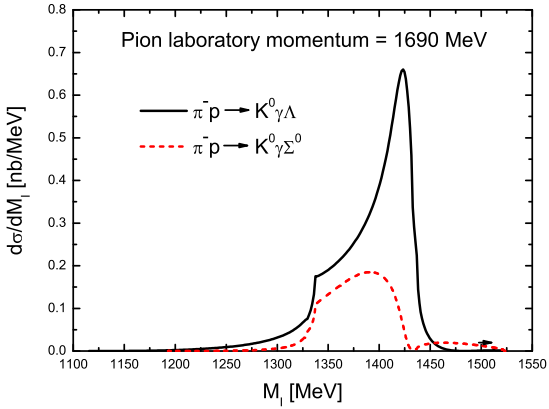}
}
\caption{The invariant mass distribution of $\pi^-p\rightarrow K^0\gamma
\Lambda(\gamma\Sigma^0)$ as a function of the invariant mass of the final 
$\gamma\Lambda$($\gamma\Sigma^0$) system.}
\label{fig:7}   
\end{center}
\end{figure}

The experimental determination of the strength and shape of these reactions is
again one good test for the two pole structure of the $\Lambda(1405)$ and its
nature as a dynamically generated resonance.

\section{The $pp\rightarrow p K^+\Lambda(1405)$ reaction}
A recent experiment performed by the ANKE collaboration at COSY \cite{zychor}
has also observed the $\Lambda(1405)$ in the 
$pp\rightarrow p K^+\Lambda(1405)$ reaction.  It is a challenge to reproduce
this reaction and in a recent work we have undertaken the task of evaluating
the invariant mass distribution. 

\begin{figure}[htpb]
 \centering
\includegraphics[scale=0.35]{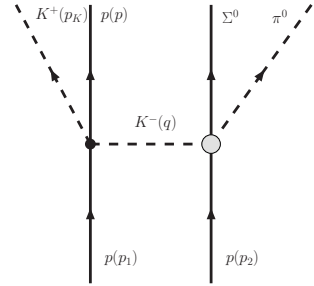}
\caption{\label{kaon_exchange} The kaon exchange mechanism of the
$pp\rightarrow pK^+\Lambda(1405)$ reaction.}
\end{figure}
\begin{figure}[htpb]
 \centering
\includegraphics[scale=0.35]{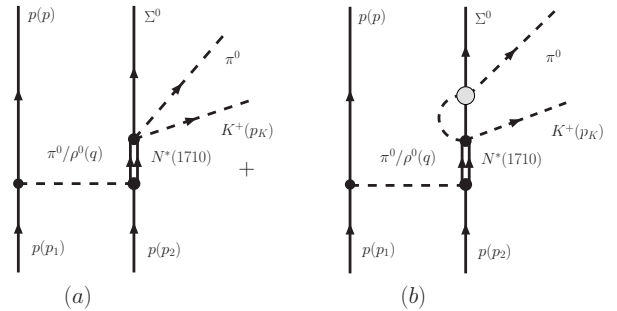}
\caption{\label{pion_exchange} The pion (rho) exchange mechanism of
the $pp\rightarrow pK^+\Lambda(1405)$ reaction through $N^*$
excitation.}
\end{figure}
\begin{figure}[htpb]
 \centering
\includegraphics[scale=0.35]{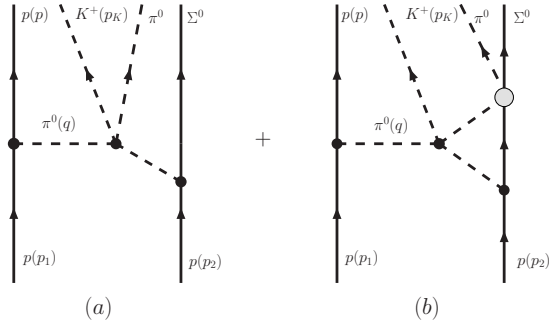}
\caption{\label{meson_pole} The pion exchange mechanism of the
$pp\rightarrow pK^+\Lambda(1405)$ reaction through meson cloud.}
\end{figure}

 The study conducted in \cite{geng} has shown that the mechanism is dominated by
 Kaon exchange in s-wave (see fig. 8), but the pion exchange (see fig. 9)
 helps bring some strength at
 lower invariant masses. The exchange of a $\rho$ meson has also been taken into
 account. Its contribution is small and contributes mostly to the uncertainties.
 The pion exchange can also contribute via the diagram of fig. 10, which
 involves the pion pion interaction. In fig. 11 we see that the shape of the cross section
 is well reproduced. We provide numbers in the figure but the comparison is made
 with 
 data which are given in arbitrary units. The integrated cross section is 
 obtained also fairly,
 but at the expense of putting some form factors in the virtual K exchange in
 s-wave, of natural size, but for which no other experimental information is
 available. The shape is rather independent on the form factor. The strength of
 the experiment, which has also large uncertainties, could be used to set some
 boundaries on the size of these form factors. 
 
 \begin{figure}[t]
 \centering
\includegraphics[scale=0.34,angle=270]{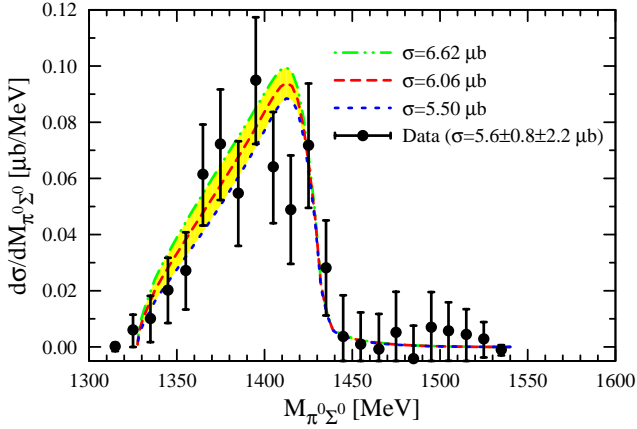}
\caption{\label{fig:error} The invariant mass distribution of the
$\pi^0\Sigma^0$ with theoretical uncertainties estimated using
Monte-Carlo sampling method (see \cite{geng} for details), in comparison
with the data~\cite{zychor}. The theoretical cross section for 
$\pi^0 \Sigma^0$ production is multiplied by a factor of three to compare with 
the experimental numbers which have also been multiplied by this factor. }
\end{figure}
 
\section{Conclusions}

We have reported on several reactions which can provide information on the
nature of the $\Lambda(1405)$ and its two pole structure, which is obtained now
in all the works using the chiral unitary approach. We have seen that in
different reactions one always has an overlap of the two resonant states in such
a way that two peaks are not seen in any reaction. Yet, the shape changes from
one to another reaction depending on the weight by which each of the states is
populated. We showed some reactions that show indeed different shapes, and
we could provide an explanation based precisely on the existence of the two
poles. Then we suggested several other reactions, feasible in present
experimental facilities, which should provide extra relevant information on the
existence of these two poles and the nature of the resonance. The predictions
made in the present work should serve to encourage the performance of these
experiments in the near future.


\end{document}